\documentclass[a4paper]{article}

\usepackage{16lomcon}        
\usepackage{cite}             
\usepackage{epsfig}           

\begin{document}


\title{RESULTS FROM THE NA61/SHINE ION PROGRAM}

\author{Tobiasz Czopowicz \email{tobiasz.czopowicz@cern.ch} for the NA61/SHINE Collaboration}

\affiliation{Faculty of Physics, Warsaw University of Technology, Warsaw, Poland}


\date{}
\maketitle


\begin{abstract}
The NA61/SHINE experiment aims to discover the critical point of strongly interacting matter
and study properties of the onset of deconfinement. These goals are to be achieved by
performing a two dimensional phase diagram ($T$-$\mu_{B}$) scan - measurements of hadron production
properties in proton-proton, proton-nucleus and nucleus-nucleus interactions as a function of
collision energy and system size.
This contribution summarizes current the status and future plans as well as presents the first physics
results of the NA61/SHINE ion program.
\end{abstract}

\section{The NA61/SHINE experiment}

NA61/SHINE ({\bf S}PS {\bf H}eavy {\bf I}on and {\bf N}eutrino {\bf E}xperiment) \cite{proposal} is
a fixed-target experiment located in the North Area of the Super Proton Synchrotron (SPS) accelerator
facility at the European Organization for Nuclear Research (CERN) in Geneva, Switzerland.
It is the successor of the NA49 experiment \cite{Afanasiev:1999iu}, which was operating in 1994 -- 2002.
The NA61/SHINE collaboration consists of 140 physicists from 28 institutes in 13 countries.
NA61/SHINE heavy-ion physics goals are:
\begin{itemize}
	\item search for the critical point of strongly interacting matter,
	\item detailed study of the onset of deconfinement,
	\item study of high $p_T$ hadron production in p+p and p+Pb interactions,
\end{itemize}

\begin{figure}[h]
	\centering
	\includegraphics[width=0.35\textwidth]{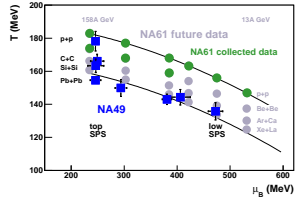}
	\caption{Freeze-out points of the planned NA61/SHINE two-dimensional phase diagram scan (based on \cite{Becattini:2005xt}).
          Be+Be collisions were recorded in 2011 and 2013.}
	\label{fig:NA61_phase_diagram_scan}
\end{figure}

In order to search for the critical point (CP) and study the properties of the onset of deconfinement, NA61/SHINE performs a two-dimensional phase diagram scan.
It measures hadron production in various collisions (p+p, Be+Be, Ar+Ca, Xe+La) at various beam energies (13$A$ -- 158$A$~GeV) \cite{add-5}.
These new data, together with Pb+Pb reactions recorded by NA49 will allow to cover the region, where the CP is expected (Fig.~\ref{fig:NA61_phase_diagram_scan}).
NA61/SHINE will search for the onset of ``kink'', ``horn'', ``step'' signatures \cite{Gazdzicki:1998vd, arXiv:0710.0118} in light nuclei
and a maximum of fluctuation signals for systems freezing-out close to the CP.

\section{Spectra and yields in p+p collisions}

\begin{figure}[ht]
	\centering
	\includegraphics[width=.35\textwidth]{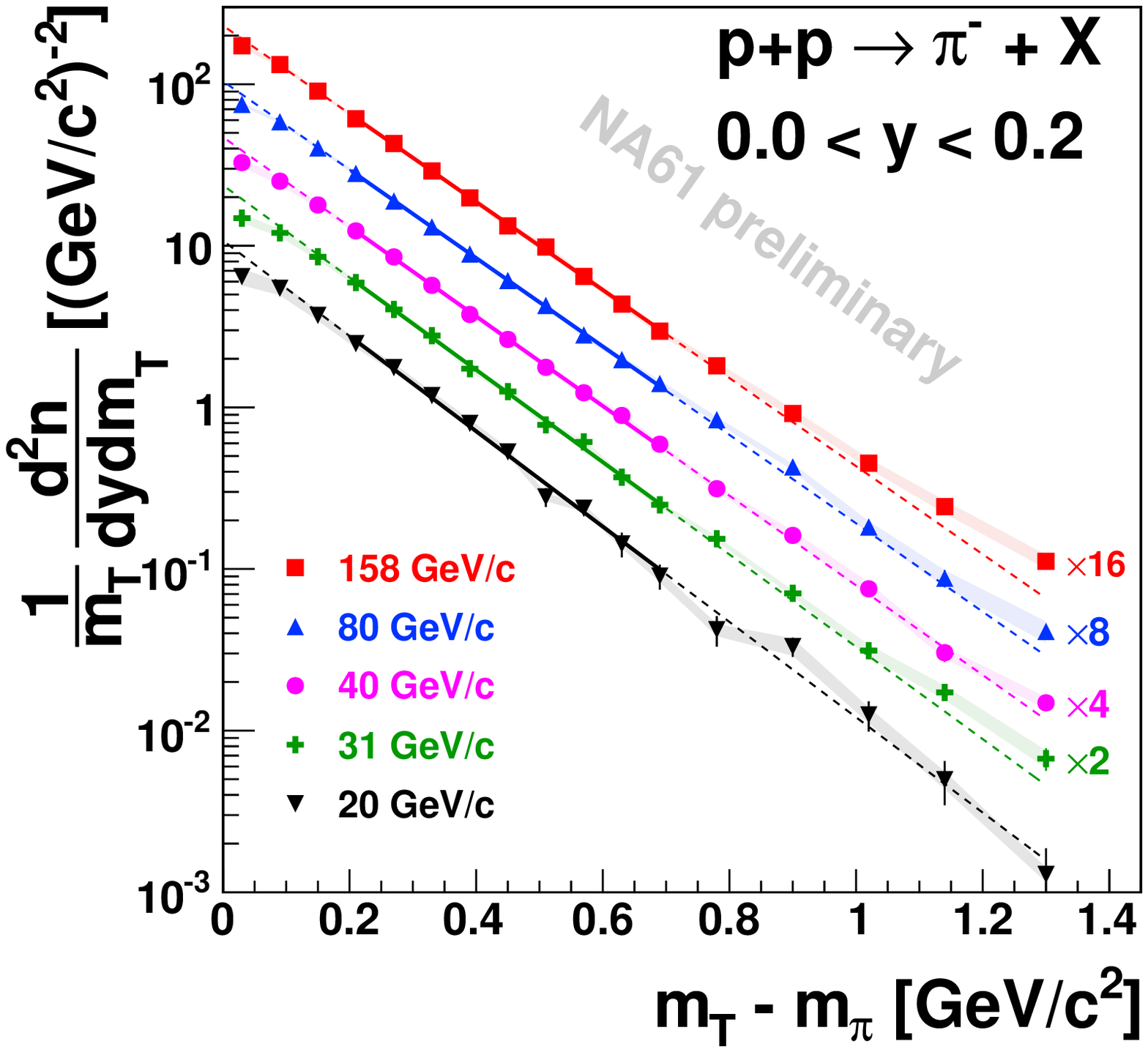} \quad
	\includegraphics[width=.35\textwidth]{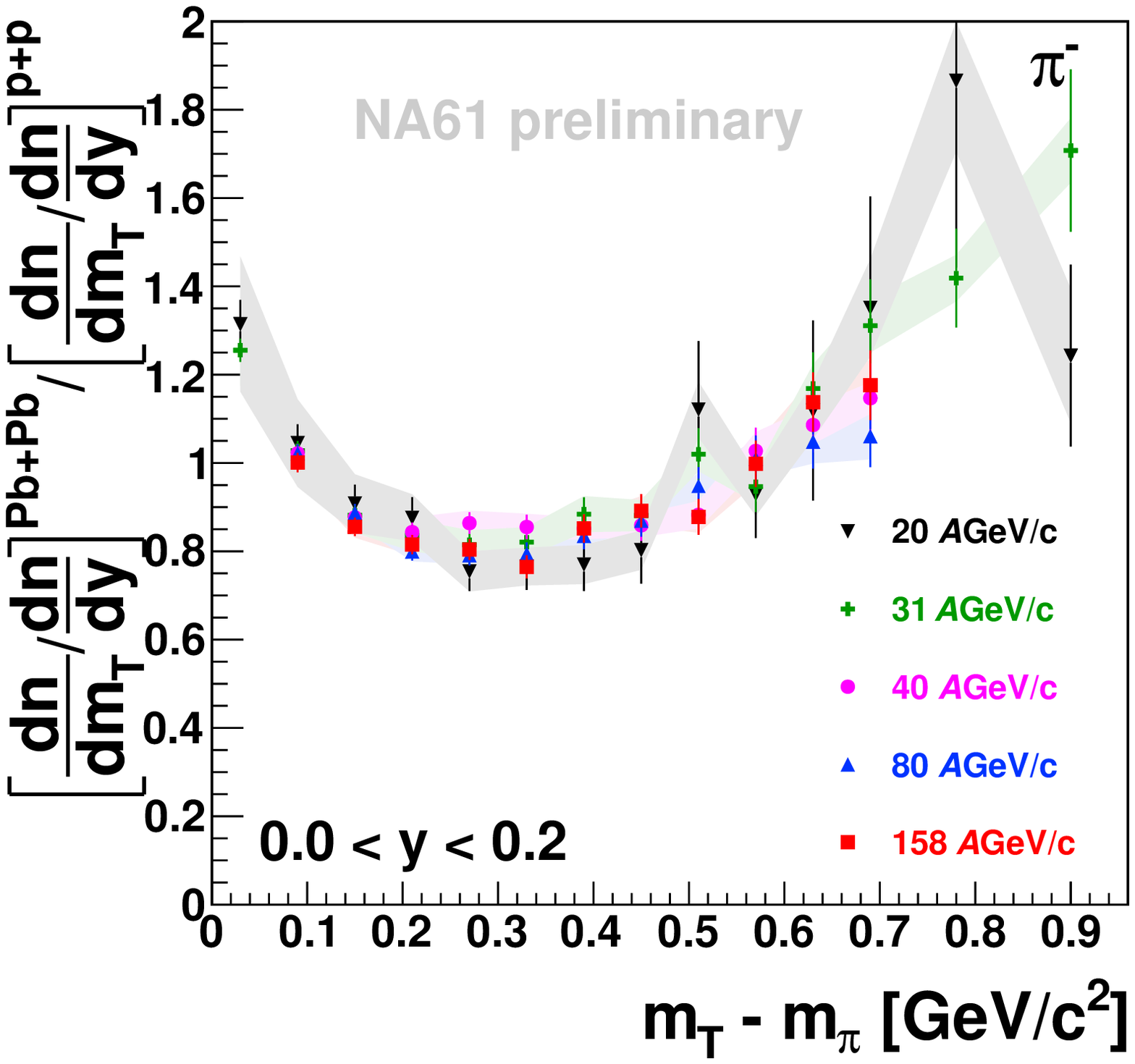} \\
	\includegraphics[width=.29\textwidth]{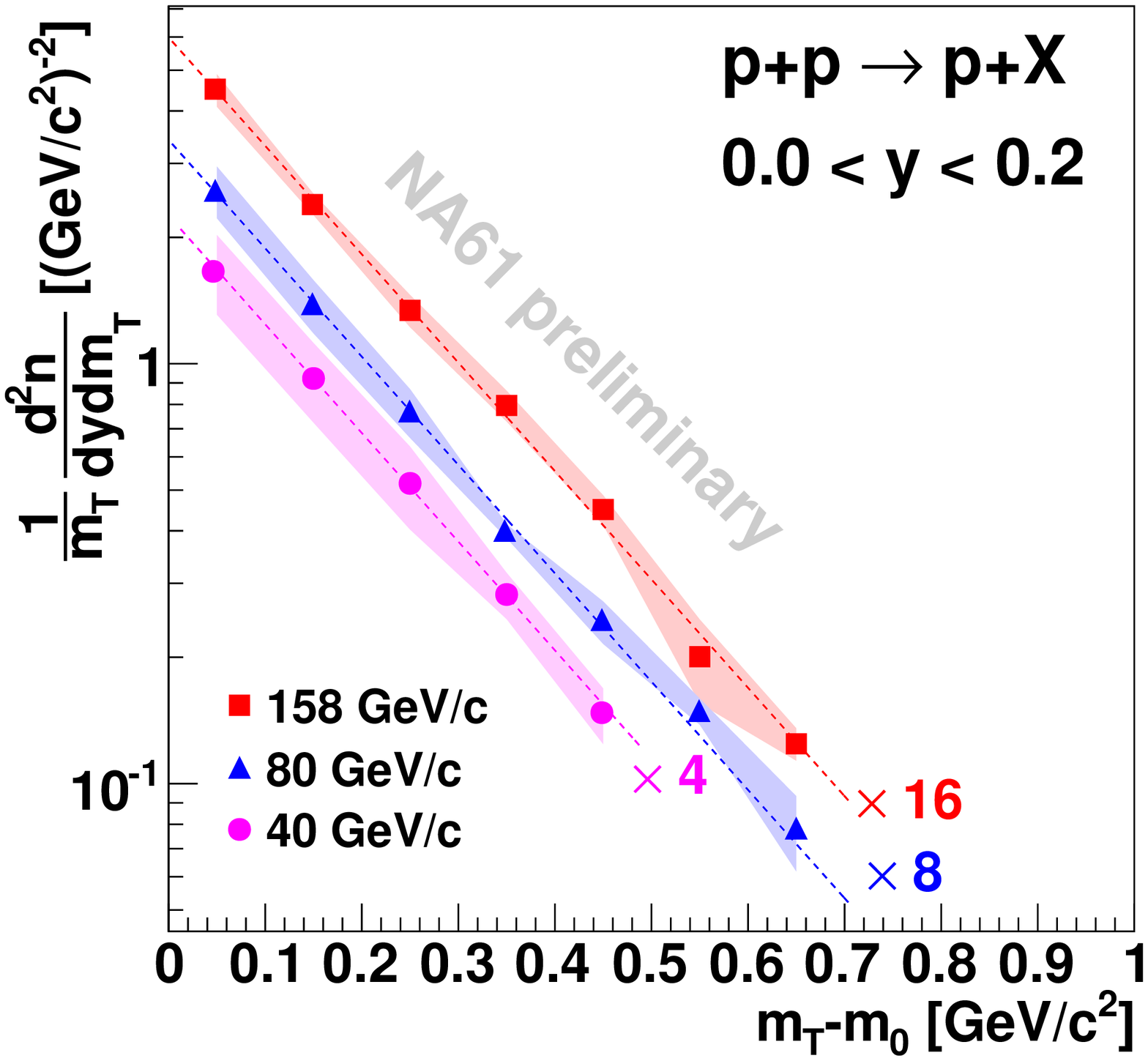} \quad\quad\quad
	\includegraphics[width=.29\textwidth]{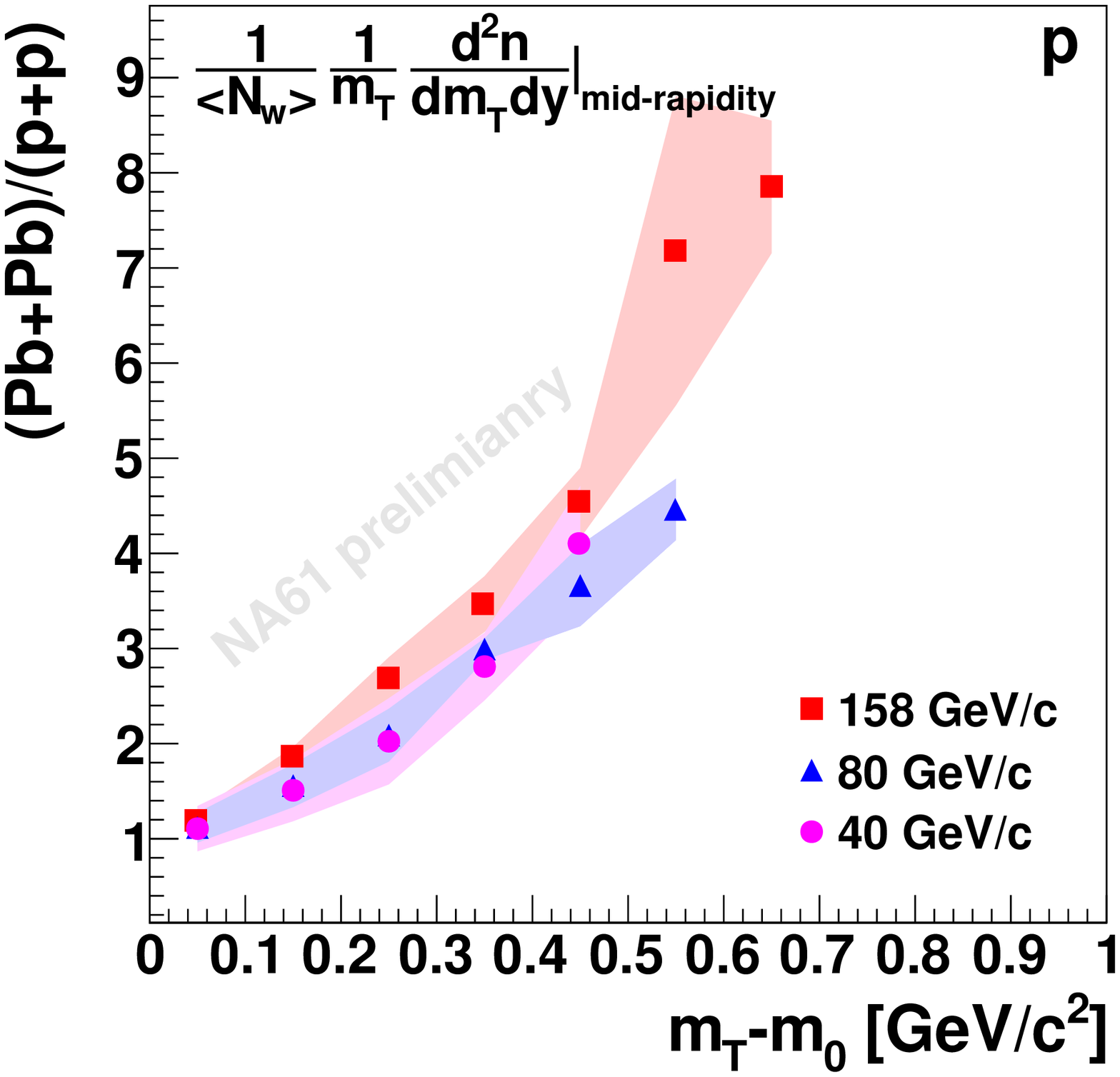}
	\caption{
		Transverse mass spectra in inelastic p+p reaction (left) and comparison with NA49 data
                from 7\% most central Pb+Pb collisions (right) for $\pi^{-}$ (top) and protons (bottom).
	}
	\label{fig:piminus}
\end{figure}

Negative pions transverse mass spectra from inelastic p+p collisions and a comparison
with NA49 data from 7\% central Pb+Pb interactions \cite{na49piminus} are shown in Fig.~\ref{fig:piminus} (top).
The results are corrected for particles from weak decays (feed-down) and detector effects using simulations.
Out of target interactions are subtracted using events recorded with empty liquid hydrogen target.
The shape of spectra differs significantly between p+p and central Pb+Pb collisions due to transverse collective
flow in Pb+Pb (see Fig.~\ref{fig:piminus} (top, right)).
No change with collision energy is observed.

Proton transverse mass spectra are shown in Fig.~\ref{fig:piminus} (bottom, left).
Comparison with NA49 data from 7\% central Pb+Pb interactions \cite{na49piminus} again show significant
shape differences due to the transverse collective flow in Pb+Pb (see Fig.~\ref{fig:piminus} (bottom, right)).
In first approximation the shape differences are independent of collision energy.

\section{Critical point and onset of deconfinement}

\begin{figure}[ht]
	\centering
	\includegraphics[width=.30\textwidth]{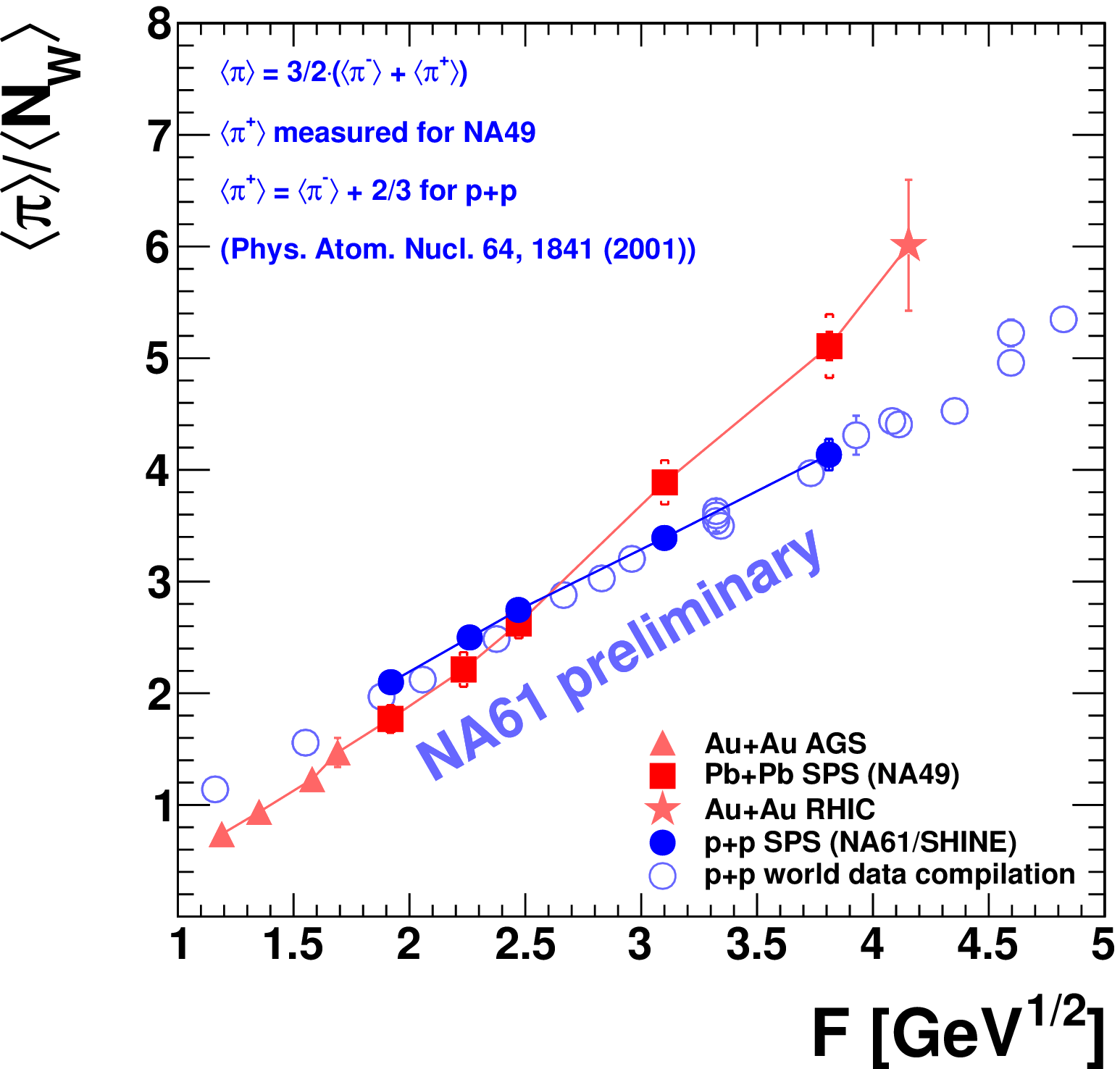} \quad
	\includegraphics[width=.30\textwidth]{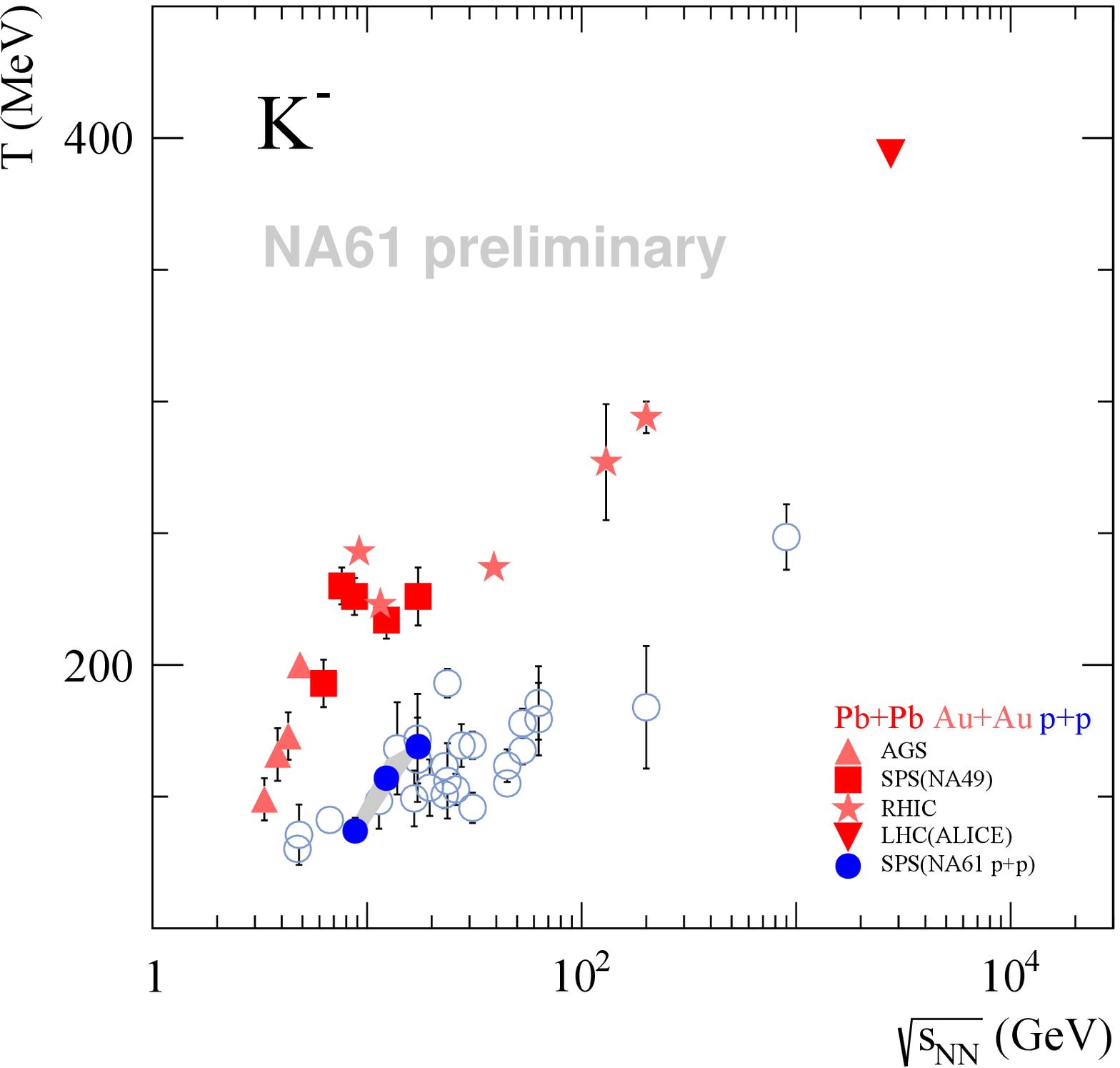} \quad
	\includegraphics[width=.32\textwidth]{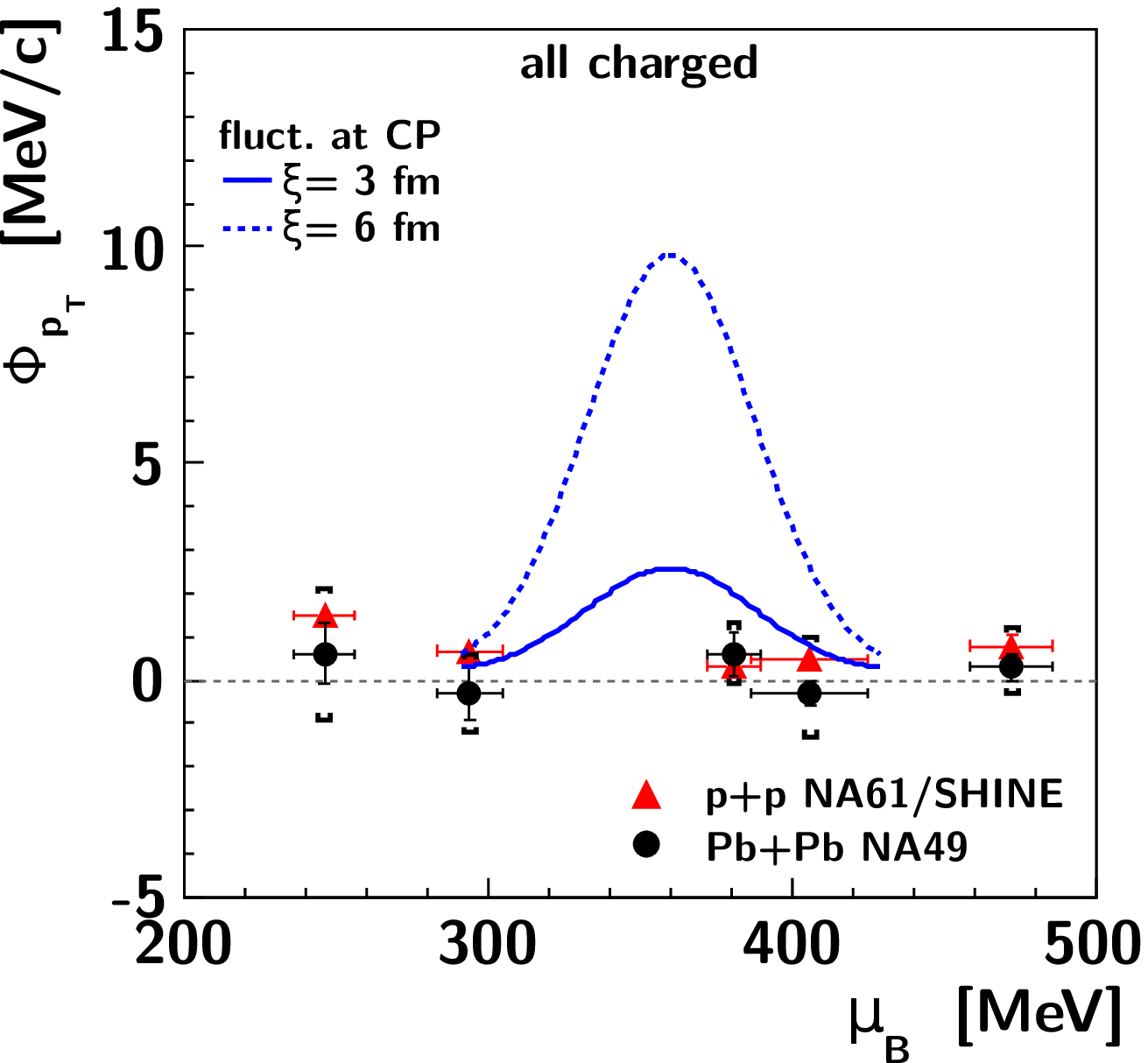}
	\caption{
		Mean $\pi^{-}$ multiplicity showing the ``kink'' structure (left),
		inverse slope parameters revealing the ``step'' structure (middle)
		and comparison of transverse momentum fluctuations with CP predictions
                for two different correlation lengths (right) in inelastic p+p (NA61/SHINE)
                and 7\% most central Pb+Pb (NA49) collisions.
	}
	\label{fig:cpod}
\end{figure}

The $\pi^{-}$ multiplicity at the SPS energies increases faster in central Pb+Pb than in p+p
collisions (``kink'', Fig.~\ref{fig:cpod} left). The two dependencies cross at about 40A GeV.
The inverse slope parameters T of $m_{T}$ spectra at the SPS energies show a different behavior in
central Pb+Pb (``step'', Fig.~\ref{fig:cpod} middle) than in p+p (smooth increase) reactions.

A comparison of transverse momentum fluctuations measured with the strongly intensive quantity $\Phi$ \cite{Phi_measure}
in NA61/SHINE inelastic p+p and NA49 7\% most central Pb+Pb collisions with CP predictions assuming
two correlation lengths is shown in the right panel of Fig.~\ref{fig:cpod}.
No indications of the CP are observed in p+p and Pb+Pb reactions. NA61/SHINE will continue to search for indications
in Be+Be, Ar+Ca and Xe+La collisions.

\section*{Acknowledgments}

This work was supported by
the Hungarian Scientific Research Fund (grants OTKA 68506 and 71989),
the Polish Ministry of Science and Higher Education (grants
667/N-CERN/2010/0, NN 202 48 4339 and  NN 202 23 1837),
the National Science Center of Poland (grant UMO-2012/04/M/ST2/00816),
the Federal Agency of Education of the Ministry of Education and Science
of the Russian Federation (grant RNP 2.2.2.2.1547), the Russian Academy of
Science and
the Russian Foundation for Basic Research (grants 08-02-00018, 09-02-00664,
and 12-02-91503-CERN),
the Ministry of Education, Culture, Sports, Science and Technology,
Japan, Grant-in-Aid for Scientific Research (grants 18071005, 19034011,
19740162, 20740160 and 20039012),
the German Research Foundation (grants GA 1480/2-1, GA 1480/2-2),
Bulgarian National Scientific Fondation (grant DDVU 02/19/ 2010),
Ministry of Education and Science of the Republic of Serbia (grant OI171002),
Swiss Nationalfonds Foundation (grant 200020-117913/1)
and ETH Research Grant TH-01 07-3.



\begin{thebibliography}{99}

\bibitem{proposal}
N.~Antoniou {\it et al.} [NA61/SHINE Collab.],
CERN-SPSC-2006-034, SPSC-P-330, (2006).

\bibitem{Afanasiev:1999iu}
S.~Afanasev {\it et al.} [NA49 Collab.],
Nucl.\ Instrum.\ Meth.\ A {\bf 430}, 210 (1999).

\bibitem{add-5}
N.~Abgrall {\it et al.} [NA61/SHINE Collab.],
CERN-SPSC-2009-031; SPSC-P-330-ADD-5.


\bibitem{Becattini:2005xt}
F.~Becattini, J.~Manninen and M.~Gazdzicki,
Phys.\ Rev.\ C {\bf 73}, 044905 (2006).

\bibitem{Gazdzicki:1998vd}
M.~Ga\'{z}dzicki, M.~Gorenstein,
Acta Phys.\ Polon.\ B {\bf 30} 2705.

\bibitem{arXiv:0710.0118}
C.~Alt {\it et al.} [NA49 Collab.],
Phys.\ Rev.\ C\ {\bf 77}, 024903  (2008).
\\
S.~V.~Afanasiev {\it et al.} [The NA49 Collab.],
Phys.\ Rev.\ C\ {\bf 66}, 054902  (2002).

\bibitem{na49piminus}
Phys.\ Rev.\ C {\bf 66}, 054902 (2002)
\\
Phys.\ Rev.\ C {\bf 77}, 024903 (2008)

\bibitem{na49pmt}
Phys.\ Rev.\ C {\bf 73}, 044910 (2006)







\bibitem{Phi_measure}
M.~Gazdzicki, S.~Mrowczynski,
Z.\ Phys.\ C {\bf 54} (1992) 127.

\end{thebibliography}
\end{document}